\documentclass[twocolumn,aps,amsmath,amssymb,prl]{revtex4-2}
\usepackage{bm}
\usepackage{xcolor}
\usepackage{graphicx}
\usepackage{boxedminipage}
\allowdisplaybreaks[4]

\newcommand{\ud}{{\mathrm d}}

\newcommand{\E}{{\mbox{\tiny E}}}
\newcommand{\tS}{{\mbox{\tiny S}}}
\newcommand{\tP}{{\mbox{\tiny P}}}

\newcommand{\D}{{\mbox{\tiny D}}}

\newcommand{\nl}{\nonumber \\}

\newcommand{\be}{\begin{equation}}
\newcommand{\ee}{\end{equation}}
\newcommand{\bea}{\begin{eqnarray}}
\newcommand{\eea}{\end{eqnarray}}
\newcommand{\bsube}{\begin{subequations}}
\newcommand{\esube}{\end{subequations}}
\newcommand{\Eq}[1]{Eq.\,(\ref{#1})}

\newcommand{\Fig}[1]{Fig.\,\ref{#1}}

\renewcommand{\ud}{\mathrm{d}}
\newcommand{\comments}[1]{}

\allowdisplaybreaks[1]

\usepackage{hyperref}  
\hypersetup{hidelinks,
	colorlinks=true,
	allcolors=black,
	pdfstartview=Fit,
	breaklinks=true
}

\begin{document}
\title{Open Quantum Systems from Dynamical Constraints}

\author{Yu Su} 
\email{suyupilemao@mail.ustc.edu.cn}
\author{Yao Wang} 
\email{wy2010@ustc.edu.cn}
\affiliation{Hefei National Research Center for Physical Sciences at the Microscale, University of Science and Technology of China, Hefei, Anhui 230026, China}

\date{\today}

\begin{abstract}
Open quantum systems are traditionally described by decomposing the total Hilbert space into a system and an external environment, linked by an explicit interaction Hamiltonian. We propose an alternative framework in which the environment is not introduced as an independent sector a priori, but instead emerges from the dynamical activation of constraints in an initially constrained quantum system. Within Dirac quantization, the physical degrees of freedom define the system, whereas the constraint sector, once promoted to carry its own dynamics, functions as an environment. In this picture, the system-environment coupling is not added through a separate interaction term, but is encoded directly in the constraint structure. As an example, we study a quantum particle coupled to a Brownian-oscillator environment and show how the resulting environmental influence can be formulated in this constraint-based setting. Our results provide a new perspective on the origin of quantum environments and point toward a general framework for open quantum systems rooted in constrained quantization.
\end{abstract}

\maketitle

\section{Introduction}
Open quantum systems are ubiquitous in various fields of
science \cite{Wei21,Kle09,Bre02,Riv12},
covering quantum optics \cite{Scu97, Lou73,Haa7398,Hak70,Sar74},
nuclear magnetic resonance \cite{Rei82,Sli90,Van051037},
condensed matter and material physics \cite{Bor85,Hol59325,Hol59343,Kli97,Ram98},
quark-gluon plasma  \cite{Aka15056002,Bla181,Miu20034011,Yao212130010},
nonlinear spectroscopy \cite{Muk95,She84,Muk81509,Yan885160,Yan91179,Che964565,Tan939496,Tan943049},
chemical and biological physics \cite{Nit06,Lee071462,Eng07782,Dor132746,Cre13253601,Kun22015101}.
Theoretically, open quantum systems are usually formulated by embedding the degrees of freedom of interest into a larger composite and then tracing over the remaining variables.
In this standard picture one begins with an a priori decomposition of the total Hilbert space into a system sector (S) and an environment sector (E), together with a Hamiltonian of the form 
\be \label{Htot_origin}
H_{\rm tot}=H_{\tS}+H_{\E}+H_{\rm int}.
\ee
The last term accounts for the interaction between the system and its environment.
This viewpoint underlies several of the most prevalent formalisms in the subject, including projection-operator techniques \cite{Nak58948,Zwa601338}, Redfield-type equations \cite{Red5719,Dav7491,Bre02}, and the theory of completely positive quantum dynamical semigroups \cite{Gor76821,Lin76119}. It also provides the natural language for modern discussions of non-Markovianity, memory effects, dissipation, and decoherence across various physical
regimes \cite{Bre09210401,Riv10050403,
Riv14094001,Bre16021002,Dev17015001,Hor09221}.

Among microscopic realizations of this program, the harmonic-oscillator bath occupies a particularly distinguished position. Rooted in the Feynman–Vernon influence functional \cite{Fey63118} and developed into the Caldeira–Leggett description of quantum dissipation, the oscillator-bath model has become the canonical framework for quantum Brownian motion, macroscopic quantum tunneling, and dissipative two-state dynamics \cite{Cal81211,Cal83587,Cal83374,Haa852462,Leg871,Gra88115,Han05026106}. In this class of models the environment is represented by a set of harmonic modes linearly coupled to the distinguished coordinates, allowing one to derive effective dissipation and noise kernels in a form that is both physically transparent and technically tractable. 
The same line of work led to a large literature on the damped quantum oscillator, exact and near-exact master equations, and the thermodynamic and dynamical subtleties of strong damping and nonlocal dissipation \cite{Xu26xxx}.

\begin{figure*}[t]
    \centering
\includegraphics[width=0.618\linewidth]{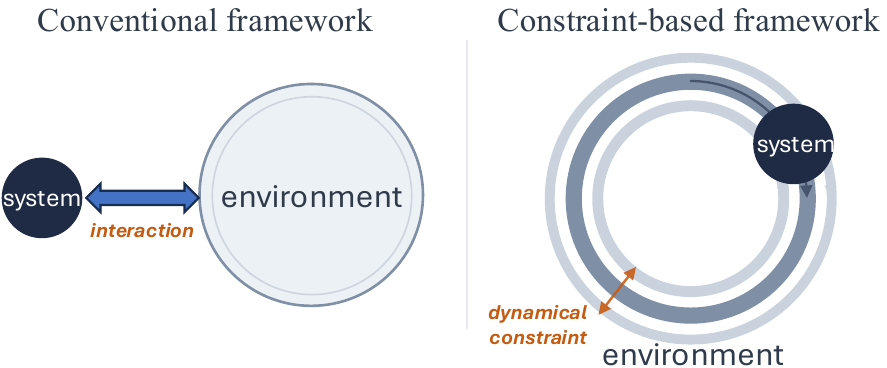}
\caption{Comparison between the conventional construction of an open quantum system and the constraint-based construction proposed in this work. In the conventional picture, the system and the environment are introduced as distinct sectors from the outset and are coupled through an explicit interaction (left panel). In the present framework, the same system is constrained to move on a ring, while the constraint sector acquires its own dynamics and thereby plays the role of an environment. The multiple ring radii indicate the breathing motion of the constraint, namely its dynamical activation. Open-system behavior then arises through coupling mediated by the dynamical constraint, rather than by a separately prescribed interaction Hamiltonian (right panel).}
    \label{fig1}
\end{figure*}

Despite its success, however, the conventional open-system construction also builds in a strong structural assumption from the outset: the system and the environment are treated as independent sectors before they are coupled. Even when one allows for initial correlations, memory kernels, colored noise, or strong coupling, the conceptual starting point remains a prescribed system-bath split together with a prescribed interaction Hamiltonian. In that sense, openness is not derived but postulated. The reduced dynamics may then be highly nontrivial, but the ontology of ``system plus external environment'' is fixed at the beginning of the construction \cite{Wei21,Kle09,Bre02}. This is precisely the standpoint that we wish to reconsider here.

There exists, on the other hand, a different and equally longstanding theoretical framework in which the distinction between physical and auxiliary degrees of freedom is organized in a fundamentally different manner, namely the Hamiltonian theory of constrained systems \cite{Dir64}. In Dirac’s formulation, constraints restrict the phase space, select the physical sector, and determine the algebraic structure of observables; quantization then proceeds by implementing these constraints and constructing the corresponding physical Hilbert space. This viewpoint was subsequently systematized and extended in the modern theory of constrained and gauge systems, with detailed treatments of first- and second-class constraints, reduced phase-space methods, BRST-related structures, and covariant quantization. Closely related insights were also developed in the symplectic reduction approach of Faddeev and Jackiw \cite{Dir64,Sun82,Hen92,Git90,Fad881692}.

The central idea of the present work is to bring these two traditions together in a different order from the conventional one. Rather than starting from an externally specified environment, we begin with a single constrained quantum system. The physical degrees of freedom of the constrained theory define the system, while degrees of freedom associated with the constraint sector are allowed to acquire their own dynamics. Once dynamically activated, this sector plays the role of an effective environment. In this way, the system–environment split is not fundamental but emergent, and the coupling between system and environment is not introduced through an additional interaction Hamiltonian. Instead, it is encoded in the constraint structure itself. From this perspective, openness arises from the dynamical extension of constraints rather than from the attachment of an external bath.

This reformulation has several immediate conceptual consequences. First, it relocates the origin of dissipation and decoherence from a postulated external sector to an intrinsic enlargement of a constrained system. Second, it suggests that the exchange of information between system and environment can be reinterpreted as the buildup of correlations between physical degrees of freedom and a dynamical constraint sector. Third, it offers a natural language for situations in which the separation between genuine physical variables and auxiliary variables is itself dynamical, rather than fixed once and for all. \emph{The aim of the present paper is not to discard the powerful machinery of conventional open-system theory, but to show that open-system can be constructed from a different structural starting point, one rooted in constrained quantization.}

As a concrete illustration, in the following we formulate the dynamics of a quantum particle coupled to a Brownian-oscillator environment in this constraint-based language.
This example connects our framework directly with the familiar Caldeira–Leggett paradigm while making transparent how an effective environment can emerge from dynamical constraints. The Brownian bath is therefore retained as a benchmark, but its conceptual status is changed: rather than being an externally appended reservoir, it is reinterpreted as the dynamical manifestation of a constraint sector. In this sense, our approach provides a new route to open quantum systems—one in which the origin of the environment itself becomes part of the dynamical construction.
We will conclude our paper by outlining a potential application of this framework in molecular dynamics.

\section{Dynamical constraint}
Let us start with the description of system. Its Hamiltonian reads
\begin{align}\label{Hs}
  H_{\tS} = \frac{\bm p^2}{2m} + V(\bm x),
\end{align} 
describing a  quantum particle of mass $m$ moving in the  potential $V(\bm x)$.
The total system is subject to an additional constraint:
\begin{align}
 \bm x^2 - R^2 = 0.
\end{align}
In the standard quantum-mechanical treatment of constrained systems, 
$R$ is typically regarded as a c-number, possibly time-dependent, that carries no dynamical degrees of freedom of its own \cite{Wei15}.
We now arrive at the first key step of this paper: we promote $R$ to a genuine \emph{classical} dynamical degree of freedom $Q$:
\be 
R \longrightarrow Q,
\ee
endowed with its own equation of motion governed by the environmental Hamiltonian $H_{\E}$.

We now return to the ``Schrödinger picture'' (in classical sense), in which the problem can be fully formulated as follows.
The total system-plus-environment Hamiltonian reads
\begin{align}\label{Htot}
  H_{\rm tot} = H_{\tS} + H_{\E}.
\end{align} 
\emph{It is important to note the absence of a system-environment interaction term}, compared with that in \Eq{Htot_origin}. 
In contrast, the dynamical variables of the system and the environment are subject to a constraint:
\begin{align}\label{constraint}
  \phi={\bm x}^2 -  Q^2 = 0.
\end{align}
 In the following, we take the example that the environment takes the form of a Brownian oscillator in the Caldeira–Leggett form:
\begin{align}\label{He}
  H_{\E} = \frac{P^2}{2M}\!+\!\frac{M\Omega^2  Q^2}{2}\!+\!\sum_j\!\bigg( \frac{\tilde p_j^2}{2m_j}\!+\! \frac{m_j\omega_j^2\tilde q_j^2}{2} \bigg)\!+Q c_j\tilde  q_j.
\end{align}
This Hamiltonian describes the environment as a distinguished Brownian oscillator \((Q, P)\) of mass \(M\) and bare frequency \(\Omega\), linearly coupled to a bath of independent harmonic oscillators \((\tilde q_j,\tilde p_j)\) with masses \(m_j\) and frequencies \(\omega_j\). The first two terms, \( P^2/2M\) and \(M\Omega^2 Q^2/2\), represent the kinetic and potential energies of the Brownian mode itself. The bath is encoded in the sum over \(j\), where each oscillator contributes its own kinetic and harmonic potential energy. The interaction term, proportional to \( Q\, c_j \tilde q_j\), is bilinear in the system and bath coordinates and specifies how the Brownian oscillator exchanges energy and information with the surrounding environment through the coupling constants \(c_j\). In this way, the Hamiltonian provides a canonical model for dissipation and decoherence induced by an oscillator bath.

Let us now briefly summarize the structure of the formulation. We start from the total Hamiltonian in \Eq{Htot}, which contains no explicit interaction term. Its system part is given in \Eq{Hs}, whereas the environmental part is specified in \Eq{He}. The coupling between the system and the environment is not introduced through a separate interaction Hamiltonian; instead, it is incorporated indirectly through the constraint given in \Eq{constraint}.
In this setting, the Poisson bracket  is defined as
\begin{align}
  \{f,g\}_{\tP} \equiv \sum_a \left( \frac{\partial f}{\partial\xi_a}\frac{\partial g}{\partial p^\xi_a} - \frac{\partial f}{\partial p^\xi_a}\frac{\partial g}{\partial \xi_a}  \right)
\end{align}
with $\{\xi_a\}$ and $\{p^\xi_a\}$ being all the phase space variables ($\bm x$, $\bm p$, $Q$, $P$, $\{\tilde q_j\}$, $\{\tilde p_j\}$).

\section{Dirac--Bergmann alogrithm}
In what follows, we handle this constraint using the Dirac--Bergmann algorithm. The self-consistent condition of \Eq{constraint} reads 
\begin{align}\label{phi-c}
  \{H_{\rm tot} + \lambda\phi, \phi\}_{\tP} \approx 0,
\end{align}
where $\lambda$ is the Lagrange multiplier, and the symbol $\approx$ means the weak equality, i.e., substituting the constraints after calculating all the Poisson brackets.
The self-consistent condition, \Eq{phi-c}, gives born to the secondary constraint
\begin{align}
  \chi = \frac{2\bm x\cdot\bm p}{m} - \frac{2Q P}{M} = 0.
\end{align}
The self-consistent condition for $\chi$ would not further generate new constraints, but gives the equation for the Lagrange multiplier $\lambda$,
\begin{align}
  \lambda = -\frac{\{H_{\rm tot},\chi\}_{\tP}}{\{\phi,\chi\}_{\tP}}.
\end{align}
Thus, $\phi$ and $\chi$ are two second class constraints, upgrading the Poisson bracket to the Dirac bracket, 
\begin{align}
  \{f,g\}_{\D} = \{f,&g\}_{\tP} + \frac{mM}{4(M\bm x^2 + mQ^2)}\nl
  &\times\big( \{f,\phi\}_{\tP}\{\chi,g\}_{\tP} - \{f,\chi\}_{\tP}\{\phi,g\}_{\tP} \big).
\end{align}
As a result, for any physical observable $A(\{\xi_a\},\{p^\xi_a\})$, the equation of motion  reads
\begin{align}
  \dot A = \{A, H_{\rm tot} + \lambda\phi\}_{\tP} = \{A, H_{\rm tot}\}_{\D}.
\end{align}

\section{Dirac quantization}
For the constrained system considered here, all constraints are of second class. Therefore the Dirac quantization is implemented by promoting any classical observable $A$ to an operator $\hat A$ and replacing the Poisson bracket by the Dirac bracket,
\begin{align}
  [\hat f,\hat g] \equiv \hat f \hat g - \hat g \hat f
  = i\hbar \{f,g\}_{\mathrm D}.
\end{align}
At the operator level, the constraints become identities,
\bsube
\begin{align}
  \hat{\bm x}^{\,2} - \hat Q^2 = 0,
\end{align}
and
\begin{align}
  M \hat{\bm x}\cdot \hat{\bm p} - m \hat Q \hat P = 0.
\end{align}
\esube
Accordingly, the total Hamiltonian keeps the additive form
\bsube
\begin{align}
  \hat H_{\mathrm{tot}} = \hat H_{\tS} + \hat H_{\E},
\end{align}
with
\begin{align}
  \hat H_{\tS} = \frac{\hat{\bm p}^{\,2}}{2m} + V(\hat{\bm x}),
\end{align}
\esube
while $\hat H_{\E}$ denotes the Brownian-oscillator environment, the quantized version of \Eq{He}.

The nontrivial Dirac brackets modify the canonical algebra. The fundamental commutation relations for the system sector are
\begin{subequations}\label{xp}
    \begin{align}
      [\hat x_i,\hat x_j] &= 0,\\
      [\hat x_i,\hat p_j] &= i\hbar\left(\delta_{ij} - \frac{M}{m+M}\frac{\hat x_i \hat x_j}{\hat{\bm x}^{\,2}}\right),\\
      [\hat p_i,\hat p_j] &= -\,i\hbar\,\frac{M}{m+M}\frac{1}{\hat{\bm x}^{\,2}} \left(\hat x_i \hat p_j - \hat x_j \hat p_i\right).
    \end{align}
\end{subequations}
For the environmental collective coordinate, one has
\begin{align}
  [\hat Q,\hat Q] &= 0, \quad
  [\hat P,\hat P] = 0, \quad
  [\hat Q,\hat P] = i\hbar \frac{M}{m+M}.
\end{align}
More importantly, the system and environment no longer commute with each other:
\begin{subequations}
    \begin{align}
  [\hat Q,\hat x_i] &= 0,\\
  [\hat Q,\hat p_i] &= i\hbar\,\frac{m}{m+M}\frac{\hat x_i}{\hat{\bm x}^{\,2}}\,\hat Q,\\
  [\hat P,\hat x_i] &= -\,i\hbar\,\frac{M}{m+M}\frac{\hat x_i}{\hat Q},\\
  [\hat P,\hat p_i] &= i\hbar\,\frac{1}{(m+M)\hat{\bm x}^{\,2}}(m \hat x_i \hat P - M \hat Q \hat p_i).
\end{align}
\end{subequations}
Hence, even though $\hat H_{\mathrm{tot}}$ is written as a sum of a system part and an environment part, the constraint induces a noncommutative structure between the two sectors, and therefore
$
  [\hat H_{\tS},\hat H_{\E}] \neq 0
$.
This shows that the interaction is not introduced by an explicit coupling term in the Hamiltonian, but is instead generated by the environmental constraint itself.

\section{Auxiliary stochastic fields} Although Dirac's prescription lays the quantization foundation of systems with constraints, it is inconvenient for practical using especially when the canonical commutators, e.g. the right-hand-side of \Eq{xp}, are function of coordinates and momenta. In such cases, the ordering of operators should be imposed to ensure that the probability measure of wave functions and the hermicity of operators are well-defined. Thus, a convenient but tricky alternative is the path integral quantization for constrained systems, which is named as the Faddeev--Senjanovic (FS) method \cite{Sen76227}. For our model only with the second-class constraints, the propagator in path integral representation reads \cite{Sen76227}
\begin{align}
  \!\!U(\xi_f,t_f;\xi_i,t_i) = \!\!\!\int\limits_{\xi(t_i) = \xi_i}^{\xi(t_f) = \xi_f}\!\!\!\!\!\sqrt{\det\mathbf C}\,\mathcal D\Gamma\,\delta(\phi)\delta(\chi)e^{\frac{i}{\hbar}S_{\rm tot}[\Gamma]},
\end{align}
where $\Gamma \equiv (\{\xi_a,p_a^\xi\})$ is the point of total phase space, $\xi \equiv (\{\xi_a\})$, $S_{\rm tot}[\Gamma]$ is the action functional, and $\mathbf C_{12} \equiv \{\phi,\chi\}_{\tP} = -\mathbf C_{21} = 4(\frac{\bm x^2}{m} + \frac{Q^2}{M})$, $\mathbf C_{11} = \mathbf C_{22} = 0$. Within the functional integration language, the operator ordering is encoded into the integral measure, $\sqrt{\det\mathbf C}\,\mathcal D\Gamma$. 

\begin{figure}[t]
    \centering
\includegraphics[width=\linewidth]{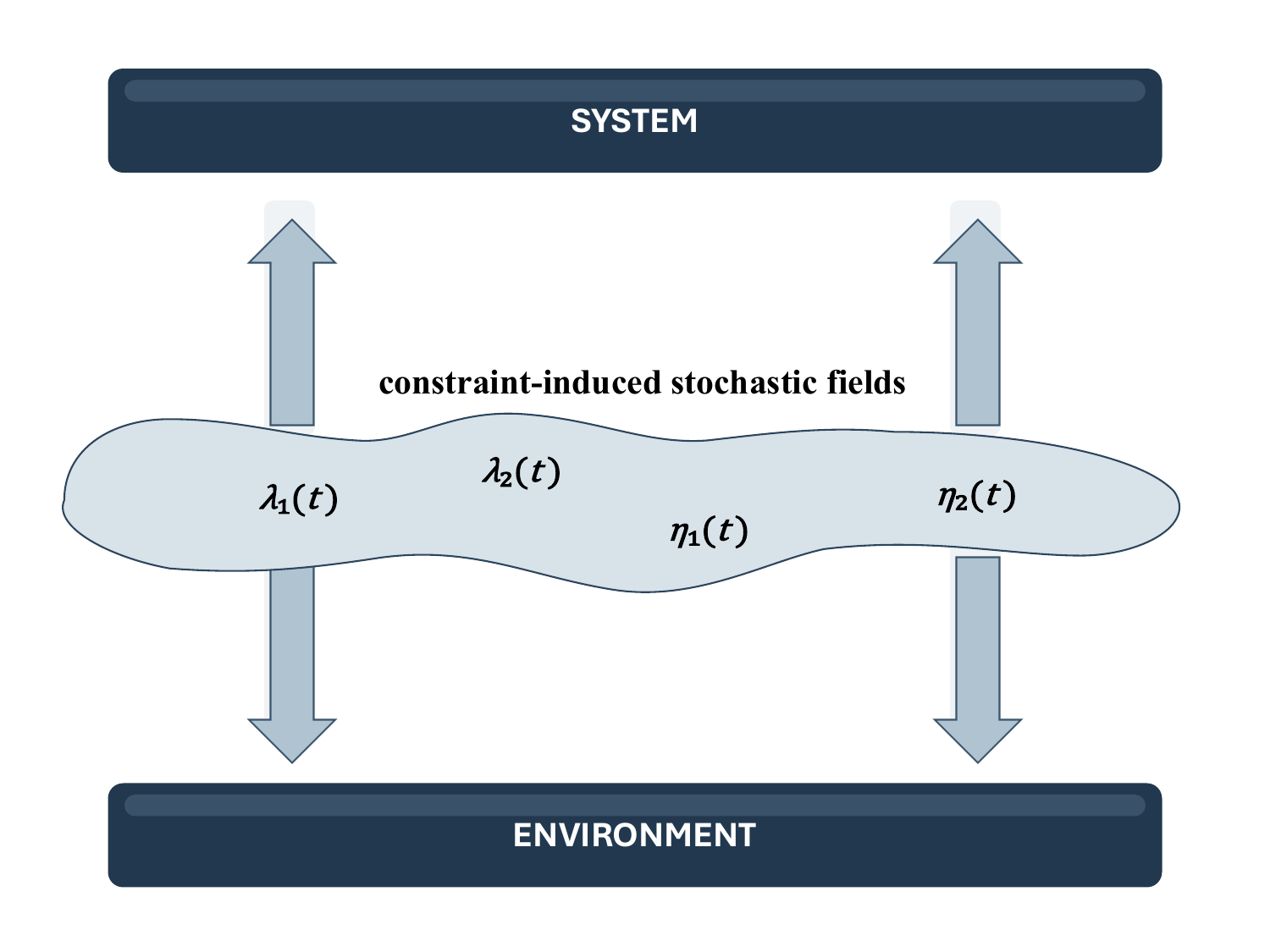}
\caption{The interaction between the system and environment is mediated with the constraint-induced stochastic fields.}
    \label{fig2}
\end{figure}

Imposing the auxiliary degrees of freedom, $\lambda = (\lambda_1,\lambda_2)$ (commutative c-number) and $\eta = (\eta_1,\eta_2)$ (anti-commutative Grassmann number), we express the propagation as 
\begin{align}
  U({\xi}_f,t_f;{\xi}_i,t_i) = \!\!\!\int\limits_{\xi(t_i) = \xi_i}^{\xi(t_f) = \xi_f}\!\!\!\mathcal D{\xi}\mathcal D\lambda\mathcal D\bar\eta\mathcal D\eta\, e^{\frac{i}{\hbar}S_{\rm tot}[{\xi},\lambda,\eta,\bar\eta]},
\end{align}
where we have used identities 
\begin{align*}
  \delta[\phi]\delta[\chi] &\propto \int\!\mathcal D\lambda\,\exp\bigg[ \frac{i}{\hbar}\int_{t_i}^{t_f}\!\!\ud t\,(\lambda_1\phi+\lambda_2\chi) \bigg], \\
  \sqrt{\det\mathbf C} &\propto \int\!\mathcal D\bar\eta\mathcal D\eta\,\exp\bigg( \frac{i}{2\hbar}\sum_{ij}\!\int_{t_i}^{t_f}\!\!\ud t\,\bar\eta_i\mathbf C_{ij}\eta_j \bigg),
\end{align*}
and integrated out the momenta. Here, the configuration space action is defined via $S_{\rm tot} = \int\!\ud t\, (\tilde L_{\tS} + \tilde L_{\E})$. The effective system and environment Lagrangians are given by 
\begin{subequations}
  \begin{align}
    \tilde L_{\tS} &= \frac{m\dot{\bm x}^2}{2} - V(\bm x) + \bigg( \lambda_1 - \dot\lambda_2 + \frac{2\lambda_2^2}{m} + \frac{2\zeta}{m} \bigg)\bm x^2,\\ 
    \tilde L_{\E} &= \frac{M\dot Q^2}{2} - \frac{M\Omega^{2}_tQ^2}{2} \nl 
    &\quad\,+ \sum_j\bigg( \frac{m_j\dot{\tilde q}_j^2}{2} - \frac{m_j\omega_j\tilde q_j^2}{2} - c_jQ\tilde q_j \bigg),
  \end{align}
\end{subequations}
where the Brownian oscillator is modified to have an effective time-dependent frequency 
\begin{align}
  \Omega^{2}_t\equiv \Omega^2 + \frac{2\lambda_1 - 2\dot\lambda_1}{M} - \frac{4\lambda_2^2}{M^2} - \frac{4\zeta}{M^2},
\end{align}
with $\zeta \equiv \bar\eta_1\eta_2 - \bar\eta_2\eta_1$. Consequently, the path integral results show that the interaction between the system and environment is mediated with the constraint-induced stochastic fields $\{\lambda(t)\}$ and $\{\eta(t)\}$ [cf.\,\Fig{fig2}]. This formalism resembles the approach developed by Shao and co-workers, in which the system-environment interaction are decoupled through stochastic fields \cite{Sha045053,Yan04216}.

\section{Summary and perspective}
The above construction reveals a new mechanism for open quantum system. In the standard
paradigm, the system couples to the environment through an explicit interaction
term added to the Hamiltonian. In contrast, in the present framework the total Hamiltonian
remains formally additive, while the constraint $\hat{\bm x}^{\,2}=\hat Q^2$ reshapes the
underlying structure and thereby induces nontrivial commutators between system and environment
variables. As a result, the environmental influence is encoded directly in the operator
algebra rather than in a separate coupling Hamiltonian.
From this perspective, the constrained construction substantially enlarges the class of
open quantum systems that can be systematically studied.

To conclude this paper, we would like to outline a potential application of the developed framework in molecular dynamics. 
A natural starting point is the Born--Oppenheimer description of molecular motion, in which the nuclear degrees of freedom evolve on an adiabatic potential-energy surface $V(\mathbf R)$ according to
$
M_A \ddot{\mathbf R}_A = - \nabla_A V(\{\mathbf R\})
$.
Although this equation defines a high-dimensional dynamics in the full nuclear configuration space, chemical reactions are often organized by a distinguished reaction path connecting reactants, transition states, and products. 
In Fukui's formulation \cite{Fuk81363}, this path is identified with the intrinsic reaction coordinate (IRC), namely the steepest-descent curve in the mass-weighted Cartesian coordinates $x_i$, for which $\ud x_i/\ud s \propto \partial V/\partial x_i$, with $\ud s^2 = \sum_i \ud x_i^2$.
Equivalently, the IRC may be written in the normalized form
\begin{equation}
\frac{\ud x_i}{\ud s} = \frac{\partial V/\partial x_i}{\ud V/\ud s},
\end{equation}
which makes explicit that the tangent vector to the path is fixed by the local gradient of the potential. 
The physical meaning of this construction is that, near a reaction event, the complicated multidimensional nuclear motion is effectively reduced to a dominant slow direction, while the remaining modes act as fast transverse fluctuations around it. 
From this perspective, the framework developed in this work suggests a possible route toward a generalized theory of reaction-path dynamics, in which environmental or collective effects may reshape the effective path itself by treating it as a dynamical constraint.
This viewpoint may be particularly valuable for molecular processes in condensed phases, where dissipation and fluctuations are expected to reorganize passage through the transition-state region in ways that are closely related to the reaction-path Hamiltonian formulation developed by Miller and co-workers \cite{Mil8099}.

Support from the National Natural Science Foundation of China (Grant Nos.\  224B2305 and 22373091) is gratefully acknowledged.


\begin{thebibliography}{67}%
\makeatletter
\providecommand \@ifxundefined [1]{%
 \@ifx{#1\undefined}
}%
\providecommand \@ifnum [1]{%
 \ifnum #1\expandafter \@firstoftwo
 \else \expandafter \@secondoftwo
 \fi
}%
\providecommand \@ifx [1]{%
 \ifx #1\expandafter \@firstoftwo
 \else \expandafter \@secondoftwo
 \fi
}%
\providecommand \natexlab [1]{#1}%
\providecommand \enquote  [1]{``#1''}%
\providecommand \bibnamefont  [1]{#1}%
\providecommand \bibfnamefont [1]{#1}%
\providecommand \citenamefont [1]{#1}%
\providecommand \href@noop [0]{\@secondoftwo}%
\providecommand \href [0]{\begingroup \@sanitize@url \@href}%
\providecommand \@href[1]{\@@startlink{#1}\@@href}%
\providecommand \@@href[1]{\endgroup#1\@@endlink}%
\providecommand \@sanitize@url [0]{\catcode `\\12\catcode `\$12\catcode
  `\&12\catcode `\#12\catcode `\^12\catcode `\_12\catcode `\%12\relax}%
\providecommand \@@startlink[1]{}%
\providecommand \@@endlink[0]{}%
\providecommand \url  [0]{\begingroup\@sanitize@url \@url }%
\providecommand \@url [1]{\endgroup\@href {#1}{\urlprefix }}%
\providecommand \urlprefix  [0]{URL }%
\providecommand \Eprint [0]{\href }%
\providecommand \doibase [0]{https://doi.org/}%
\providecommand \selectlanguage [0]{\@gobble}%
\providecommand \bibinfo  [0]{\@secondoftwo}%
\providecommand \bibfield  [0]{\@secondoftwo}%
\providecommand \translation [1]{[#1]}%
\providecommand \BibitemOpen [0]{}%
\providecommand \bibitemStop [0]{}%
\providecommand \bibitemNoStop [0]{.\EOS\space}%
\providecommand \EOS [0]{\spacefactor3000\relax}%
\providecommand \BibitemShut  [1]{\csname bibitem#1\endcsname}%
\let\auto@bib@innerbib\@empty
\bibitem [{\citenamefont {Weiss}(2021)}]{Wei21}%
  \BibitemOpen
  \bibfield  {author} {\bibinfo {author} {\bibfnamefont {U.}~\bibnamefont
  {Weiss}},\ }\href@noop {} {\emph {\bibinfo {title} {Quantum Dissipative
  Systems}}}\ (\bibinfo  {publisher} {World Scientific},\ \bibinfo {address}
  {Singapore},\ \bibinfo {year} {2021})\ \bibinfo {note} {5th
  edition}\BibitemShut {NoStop}%
\bibitem [{\citenamefont {Kleinert}(2009)}]{Kle09}%
  \BibitemOpen
  \bibfield  {author} {\bibinfo {author} {\bibfnamefont {H.}~\bibnamefont
  {Kleinert}},\ }\href@noop {} {\emph {\bibinfo {title} {Path Integrals in
  Quantum Mechanics, Statistics, Polymer Physics, and Financial Markets}}},\
  \bibinfo {edition} {5th}\ ed.\ (\bibinfo  {publisher} {World Scientific},\
  \bibinfo {address} {Singapore},\ \bibinfo {year} {2009})\BibitemShut
  {NoStop}%
\bibitem [{\citenamefont {Breuer}\ and\ \citenamefont
  {Petruccione}(2002)}]{Bre02}%
  \BibitemOpen
  \bibfield  {author} {\bibinfo {author} {\bibfnamefont {H.~P.}\ \bibnamefont
  {Breuer}}\ and\ \bibinfo {author} {\bibfnamefont {F.}~\bibnamefont
  {Petruccione}},\ }\href@noop {} {\emph {\bibinfo {title} {The Theory of Open
  Quantum Systems}}}\ (\bibinfo  {publisher} {Oxford University Press},\
  \bibinfo {address} {New York},\ \bibinfo {year} {2002})\BibitemShut {NoStop}%
\bibitem [{\citenamefont {Rivas}\ and\ \citenamefont {Huelga}(2012)}]{Riv12}%
  \BibitemOpen
  \bibfield  {author} {\bibinfo {author} {\bibfnamefont {{\'A}.}~\bibnamefont
  {Rivas}}\ and\ \bibinfo {author} {\bibfnamefont {S.~F.}\ \bibnamefont
  {Huelga}},\ }\href {https://doi.org/10.1007/978-3-642-23354-8} {\emph
  {\bibinfo {title} {Open Quantum Systems: An Introduction}}}\ (\bibinfo
  {publisher} {Springer},\ \bibinfo {address} {Heidelberg},\ \bibinfo {year}
  {2012})\BibitemShut {NoStop}%
\bibitem [{\citenamefont {Scully}\ and\ \citenamefont {Zubairy}(1997)}]{Scu97}%
  \BibitemOpen
  \bibfield  {author} {\bibinfo {author} {\bibfnamefont {M.~O.}\ \bibnamefont
  {Scully}}\ and\ \bibinfo {author} {\bibfnamefont {M.~S.}\ \bibnamefont
  {Zubairy}},\ }\href@noop {} {\emph {\bibinfo {title} {Quantum Optics}}}\
  (\bibinfo  {publisher} {Cambridge University Press},\ \bibinfo {address}
  {Cambridge},\ \bibinfo {year} {1997})\BibitemShut {NoStop}%
\bibitem [{\citenamefont {Louisell}(1973)}]{Lou73}%
  \BibitemOpen
  \bibfield  {author} {\bibinfo {author} {\bibfnamefont {W.~H.}\ \bibnamefont
  {Louisell}},\ }\href@noop {} {\emph {\bibinfo {title} {Quantum Statistical
  Properties of Radiation}}}\ (\bibinfo  {publisher} {Wiley},\ \bibinfo
  {address} {New York},\ \bibinfo {year} {1973})\BibitemShut {NoStop}%
\bibitem [{\citenamefont {Haake}(1973)}]{Haa7398}%
  \BibitemOpen
  \bibfield  {author} {\bibinfo {author} {\bibfnamefont {F.}~\bibnamefont
  {Haake}},\ }\bibfield  {title} {\bibinfo {title} {Statistical treatment of
  open systems by generalized master equations},\ }in\ \href@noop {} {\emph
  {\bibinfo {booktitle} {Quantum Statistics in Optics and Solid State Physics:
  Springer Tracts in Modern Physics, Vol.~66}}},\ \bibinfo {editor} {edited by\
  \bibinfo {editor} {\bibfnamefont {G.}~\bibnamefont {{H\"{o}hler}}}}\
  (\bibinfo  {publisher} {Springer},\ \bibinfo {address} {Berlin},\ \bibinfo
  {year} {1973})\ pp.\ \bibinfo {pages} {98--168}\BibitemShut {NoStop}%
\bibitem [{\citenamefont {Haken}(1970)}]{Hak70}%
  \BibitemOpen
  \bibfield  {author} {\bibinfo {author} {\bibfnamefont {H.}~\bibnamefont
  {Haken}},\ }\href@noop {} {\emph {\bibinfo {title} {Laser Theory}}}\
  (\bibinfo  {publisher} {Springer},\ \bibinfo {address} {Berlin},\ \bibinfo
  {year} {1970})\BibitemShut {NoStop}%
\bibitem [{\citenamefont {{Sargent~III}}\ \emph {et~al.}(1974)\citenamefont
  {{Sargent~III}}, \citenamefont {Scully},\ and\ \citenamefont
  {W.~E.~Lamb}}]{Sar74}%
  \BibitemOpen
  \bibfield  {author} {\bibinfo {author} {\bibfnamefont {M.}~\bibnamefont
  {{Sargent~III}}}, \bibinfo {author} {\bibfnamefont {M.~O.}\ \bibnamefont
  {Scully}},\ and\ \bibinfo {author} {\bibfnamefont {J.}~\bibnamefont
  {W.~E.~Lamb}},\ }\href@noop {} {\emph {\bibinfo {title} {Laser Physics}}}\
  (\bibinfo  {publisher} {Addison-Wesley},\ \bibinfo {address} {Reading, MA},\
  \bibinfo {year} {1974})\BibitemShut {NoStop}%
\bibitem [{\citenamefont {Reineker}(1982)}]{Rei82}%
  \BibitemOpen
  \bibfield  {author} {\bibinfo {author} {\bibfnamefont {P.}~\bibnamefont
  {Reineker}},\ }\href@noop {} {\emph {\bibinfo {title} {Exciton Dynamics in
  Molecular Crystals and Aggregates: Stochastic Liouville Equation Approach:
  Coupled Coherent and Incoherent Motion, Optical Line Shapes, Magnetic
  Resonance Phenomena}}}\ (\bibinfo  {publisher} {Springer},\ \bibinfo
  {address} {Berlin},\ \bibinfo {year} {1982})\BibitemShut {NoStop}%
\bibitem [{\citenamefont {Slichter}(1990)}]{Sli90}%
  \BibitemOpen
  \bibfield  {author} {\bibinfo {author} {\bibfnamefont {C.~P.}\ \bibnamefont
  {Slichter}},\ }\href@noop {} {\emph {\bibinfo {title} {Principles of Magnetic
  Resonance}}}\ (\bibinfo  {publisher} {Springer Verlag},\ \bibinfo {address}
  {New York},\ \bibinfo {year} {1990})\BibitemShut {NoStop}%
\bibitem [{\citenamefont {Vandersypen}\ and\ \citenamefont
  {Chuang}(2005)}]{Van051037}%
  \BibitemOpen
  \bibfield  {author} {\bibinfo {author} {\bibfnamefont {L.~M.~K.}\
  \bibnamefont {Vandersypen}}\ and\ \bibinfo {author} {\bibfnamefont {I.~L.}\
  \bibnamefont {Chuang}},\ }\bibfield  {title} {\bibinfo {title} {Nmr
  techniques for quantum control and computation},\ }\href@noop {} {\bibfield
  {journal} {\bibinfo  {journal} {Rev. Mod. Phys.}\ }\textbf {\bibinfo {volume}
  {76}},\ \bibinfo {pages} {1037} (\bibinfo {year} {2005})}\BibitemShut
  {NoStop}%
\bibitem [{\citenamefont {Born}\ and\ \citenamefont {Huang}(1985)}]{Bor85}%
  \BibitemOpen
  \bibfield  {author} {\bibinfo {author} {\bibfnamefont {M.}~\bibnamefont
  {Born}}\ and\ \bibinfo {author} {\bibfnamefont {K.}~\bibnamefont {Huang}},\
  }\href@noop {} {\emph {\bibinfo {title} {Dynamical Theory of Crystal
  Lattices}}}\ (\bibinfo  {publisher} {Oxford University Press},\ \bibinfo
  {address} {New York},\ \bibinfo {year} {1985})\BibitemShut {NoStop}%
\bibitem [{\citenamefont {Holstein}(1959{\natexlab{a}})}]{Hol59325}%
  \BibitemOpen
  \bibfield  {author} {\bibinfo {author} {\bibfnamefont {T.}~\bibnamefont
  {Holstein}},\ }\bibfield  {title} {\bibinfo {title} {Studies of polaron
  motion {Part I.\ The} molecular-crystal model},\ }\href@noop {} {\bibfield
  {journal} {\bibinfo  {journal} {Ann. Phys.}\ }\textbf {\bibinfo {volume}
  {8}},\ \bibinfo {pages} {325} (\bibinfo {year}
  {1959}{\natexlab{a}})}\BibitemShut {NoStop}%
\bibitem [{\citenamefont {Holstein}(1959{\natexlab{b}})}]{Hol59343}%
  \BibitemOpen
  \bibfield  {author} {\bibinfo {author} {\bibfnamefont {T.}~\bibnamefont
  {Holstein}},\ }\bibfield  {title} {\bibinfo {title} {Studies of polaron
  motion {Part II.\ The} ``small'' polaron},\ }\href@noop {} {\bibfield
  {journal} {\bibinfo  {journal} {Ann. Phys.}\ }\textbf {\bibinfo {volume}
  {8}},\ \bibinfo {pages} {343} (\bibinfo {year}
  {1959}{\natexlab{b}})}\BibitemShut {NoStop}%
\bibitem [{\citenamefont {Klingshirn}(1997)}]{Kli97}%
  \BibitemOpen
  \bibfield  {author} {\bibinfo {author} {\bibfnamefont {C.~F.}\ \bibnamefont
  {Klingshirn}},\ }\href@noop {} {\emph {\bibinfo {title} {Semiconductor
  Optics}}}\ (\bibinfo  {publisher} {Springer-Verlag},\ \bibinfo {address}
  {Heidelberg},\ \bibinfo {year} {1997})\BibitemShut {NoStop}%
\bibitem [{\citenamefont {Rammer}(1998)}]{Ram98}%
  \BibitemOpen
  \bibfield  {author} {\bibinfo {author} {\bibfnamefont {J.}~\bibnamefont
  {Rammer}},\ }\href@noop {} {\emph {\bibinfo {title} {Quantum Transport
  Theory}}}\ (\bibinfo  {publisher} {Perseus Books},\ \bibinfo {address}
  {Reading, Mass.},\ \bibinfo {year} {1998})\BibitemShut {NoStop}%
\bibitem [{\citenamefont {Akamatsu}(2015)}]{Aka15056002}%
  \BibitemOpen
  \bibfield  {author} {\bibinfo {author} {\bibfnamefont {Y.}~\bibnamefont
  {Akamatsu}},\ }\bibfield  {title} {\bibinfo {title} {Heavy quark master
  equations in the lindblad form at high temperatures},\ }\href@noop {}
  {\bibfield  {journal} {\bibinfo  {journal} {Phys. Rev. D}\ }\textbf {\bibinfo
  {volume} {91}},\ \bibinfo {pages} {056002} (\bibinfo {year}
  {2015})}\BibitemShut {NoStop}%
\bibitem [{\citenamefont {Blaizot}\ and\ \citenamefont
  {Escobedo}(2018)}]{Bla181}%
  \BibitemOpen
  \bibfield  {author} {\bibinfo {author} {\bibfnamefont {J.-P.}\ \bibnamefont
  {Blaizot}}\ and\ \bibinfo {author} {\bibfnamefont {M.~A.}\ \bibnamefont
  {Escobedo}},\ }\bibfield  {title} {\bibinfo {title} {Quantum and classical
  dynamics of heavy quarks in a quark-gluon plasma},\ }\href@noop {} {\bibfield
   {journal} {\bibinfo  {journal} {J. High Energy Phys.}\ }\textbf {\bibinfo
  {volume} {2018}},\ \bibinfo {pages} {1}}\BibitemShut {NoStop}%
\bibitem [{\citenamefont {Miura}\ \emph {et~al.}(2020)\citenamefont {Miura},
  \citenamefont {Akamatsu}, \citenamefont {Asakawa},\ and\ \citenamefont
  {Rothkopf}}]{Miu20034011}%
  \BibitemOpen
  \bibfield  {author} {\bibinfo {author} {\bibfnamefont {T.}~\bibnamefont
  {Miura}}, \bibinfo {author} {\bibfnamefont {Y.}~\bibnamefont {Akamatsu}},
  \bibinfo {author} {\bibfnamefont {M.}~\bibnamefont {Asakawa}},\ and\ \bibinfo
  {author} {\bibfnamefont {A.}~\bibnamefont {Rothkopf}},\ }\bibfield  {title}
  {\bibinfo {title} {Quantum brownian motion of a heavy quark pair in the
  quark-gluon plasma},\ }\href@noop {} {\bibfield  {journal} {\bibinfo
  {journal} {Phys. Rev. D}\ }\textbf {\bibinfo {volume} {101}},\ \bibinfo
  {pages} {034011} (\bibinfo {year} {2020})}\BibitemShut {NoStop}%
\bibitem [{\citenamefont {Yao}(2021)}]{Yao212130010}%
  \BibitemOpen
  \bibfield  {author} {\bibinfo {author} {\bibfnamefont {X.}~\bibnamefont
  {Yao}},\ }\bibfield  {title} {\bibinfo {title} {Open quantum systems for
  quarkonia},\ }\href@noop {} {\bibfield  {journal} {\bibinfo  {journal} {Int.
  J. Mod. Phys. A}\ }\textbf {\bibinfo {volume} {36}},\ \bibinfo {pages}
  {2130010} (\bibinfo {year} {2021})}\BibitemShut {NoStop}%
\bibitem [{\citenamefont {Mukamel}(1995)}]{Muk95}%
  \BibitemOpen
  \bibfield  {author} {\bibinfo {author} {\bibfnamefont {S.}~\bibnamefont
  {Mukamel}},\ }\href@noop {} {\emph {\bibinfo {title} {The Principles of
  Nonlinear Optical Spectroscopy}}}\ (\bibinfo  {publisher} {Oxford University
  Press},\ \bibinfo {address} {New York},\ \bibinfo {year} {1995})\BibitemShut
  {NoStop}%
\bibitem [{\citenamefont {Shen}(1984)}]{She84}%
  \BibitemOpen
  \bibfield  {author} {\bibinfo {author} {\bibfnamefont {Y.~R.}\ \bibnamefont
  {Shen}},\ }\href@noop {} {\emph {\bibinfo {title} {The Principles of
  Nonlinear Optics}}}\ (\bibinfo  {publisher} {Wiley},\ \bibinfo {address} {New
  York},\ \bibinfo {year} {1984})\BibitemShut {NoStop}%
\bibitem [{\citenamefont {Mukamel}(1981)}]{Muk81509}%
  \BibitemOpen
  \bibfield  {author} {\bibinfo {author} {\bibfnamefont {S.}~\bibnamefont
  {Mukamel}},\ }\bibfield  {title} {\bibinfo {title} {Reduced equations of
  motion for collisionless molecular multiphoton processes},\ }\href@noop {}
  {\bibfield  {journal} {\bibinfo  {journal} {Adv. Chem. Phys.}\ }\textbf
  {\bibinfo {volume} {47}},\ \bibinfo {pages} {509} (\bibinfo {year}
  {1981})}\BibitemShut {NoStop}%
\bibitem [{\citenamefont {Yan}\ and\ \citenamefont
  {Mukamel}(1988)}]{Yan885160}%
  \BibitemOpen
  \bibfield  {author} {\bibinfo {author} {\bibfnamefont {Y.~J.}\ \bibnamefont
  {Yan}}\ and\ \bibinfo {author} {\bibfnamefont {S.}~\bibnamefont {Mukamel}},\
  }\bibfield  {title} {\bibinfo {title} {Electronic dephasing, vibrational
  relaxation, and solvent friction in molecular nonlinear optical lineshapes},\
  }\href@noop {} {\bibfield  {journal} {\bibinfo  {journal} {J. Chem. Phys.}\
  }\textbf {\bibinfo {volume} {89}},\ \bibinfo {pages} {5160} (\bibinfo {year}
  {1988})}\BibitemShut {NoStop}%
\bibitem [{\citenamefont {Yan}\ and\ \citenamefont {Mukamel}(1991)}]{Yan91179}%
  \BibitemOpen
  \bibfield  {author} {\bibinfo {author} {\bibfnamefont {Y.~J.}\ \bibnamefont
  {Yan}}\ and\ \bibinfo {author} {\bibfnamefont {S.}~\bibnamefont {Mukamel}},\
  }\bibfield  {title} {\bibinfo {title} {Photon echoes of polyatomic molecules
  in condensed phases},\ }\href@noop {} {\bibfield  {journal} {\bibinfo
  {journal} {J. Chem. Phys.}\ }\textbf {\bibinfo {volume} {94}},\ \bibinfo
  {pages} {179} (\bibinfo {year} {1991})}\BibitemShut {NoStop}%
\bibitem [{\citenamefont {Chernyak}\ and\ \citenamefont
  {Mukamel}(1996)}]{Che964565}%
  \BibitemOpen
  \bibfield  {author} {\bibinfo {author} {\bibfnamefont {V.}~\bibnamefont
  {Chernyak}}\ and\ \bibinfo {author} {\bibfnamefont {S.}~\bibnamefont
  {Mukamel}},\ }\bibfield  {title} {\bibinfo {title} {Collective coordinates
  for nuclear spectral densities in energy transfer and femtosecond
  spectroscopy of molecular aggregates},\ }\href@noop {} {\bibfield  {journal}
  {\bibinfo  {journal} {J. Chem. Phys.}\ }\textbf {\bibinfo {volume} {105}},\
  \bibinfo {pages} {4565} (\bibinfo {year} {1996})}\BibitemShut {NoStop}%
\bibitem [{\citenamefont {Tanimura}\ and\ \citenamefont
  {Mukamel}(1993)}]{Tan939496}%
  \BibitemOpen
  \bibfield  {author} {\bibinfo {author} {\bibfnamefont {Y.}~\bibnamefont
  {Tanimura}}\ and\ \bibinfo {author} {\bibfnamefont {S.}~\bibnamefont
  {Mukamel}},\ }\bibfield  {title} {\bibinfo {title} {Two-dimensional
  femtosecond vibrational spectroscopy of liquids},\ }\href@noop {} {\bibfield
  {journal} {\bibinfo  {journal} {J. Chem. Phys.}\ }\textbf {\bibinfo {volume}
  {99}},\ \bibinfo {pages} {9496} (\bibinfo {year} {1993})}\BibitemShut
  {NoStop}%
\bibitem [{\citenamefont {Tanimura}\ and\ \citenamefont
  {Mukamel}(1994)}]{Tan943049}%
  \BibitemOpen
  \bibfield  {author} {\bibinfo {author} {\bibfnamefont {Y.}~\bibnamefont
  {Tanimura}}\ and\ \bibinfo {author} {\bibfnamefont {S.}~\bibnamefont
  {Mukamel}},\ }\bibfield  {title} {\bibinfo {title} {Multistate quantum
  fokker-planck approach to nonadiabatic wave packet dynamics in pump-probe
  spectroscopy},\ }\href@noop {} {\bibfield  {journal} {\bibinfo  {journal} {J.
  Chem. Phys.}\ }\textbf {\bibinfo {volume} {101}},\ \bibinfo {pages} {3049}
  (\bibinfo {year} {1994})}\BibitemShut {NoStop}%
\bibitem [{\citenamefont {Nitzan}(2006)}]{Nit06}%
  \BibitemOpen
  \bibfield  {author} {\bibinfo {author} {\bibfnamefont {A.}~\bibnamefont
  {Nitzan}},\ }\href@noop {} {\emph {\bibinfo {title} {Chemical Dynamics in
  Condensed Phases: Relaxation, Transfer and Reactions in Condensed Molecular
  Systems}}}\ (\bibinfo  {publisher} {Oxford University Press},\ \bibinfo
  {address} {New York},\ \bibinfo {year} {2006})\BibitemShut {NoStop}%
\bibitem [{\citenamefont {Lee}\ \emph {et~al.}(2007)\citenamefont {Lee},
  \citenamefont {Cheng},\ and\ \citenamefont {Fleming}}]{Lee071462}%
  \BibitemOpen
  \bibfield  {author} {\bibinfo {author} {\bibfnamefont {H.}~\bibnamefont
  {Lee}}, \bibinfo {author} {\bibfnamefont {Y.-C.}\ \bibnamefont {Cheng}},\
  and\ \bibinfo {author} {\bibfnamefont {G.~R.}\ \bibnamefont {Fleming}},\
  }\bibfield  {title} {\bibinfo {title} {Coherence dynamics in photosynthesis:
  Protein protection of excitonic coherence},\ }\href@noop {} {\bibfield
  {journal} {\bibinfo  {journal} {Science}\ }\textbf {\bibinfo {volume}
  {316}},\ \bibinfo {pages} {1462} (\bibinfo {year} {2007})}\BibitemShut
  {NoStop}%
\bibitem [{\citenamefont {Engel}\ \emph {et~al.}(2007)\citenamefont {Engel},
  \citenamefont {Calhoun}, \citenamefont {Read}, \citenamefont {Ahn},
  \citenamefont {Man\v{c}al}, \citenamefont {Cheng}, \citenamefont
  {Blankenship},\ and\ \citenamefont {Fleming}}]{Eng07782}%
  \BibitemOpen
  \bibfield  {author} {\bibinfo {author} {\bibfnamefont {G.~S.}\ \bibnamefont
  {Engel}}, \bibinfo {author} {\bibfnamefont {T.~R.}\ \bibnamefont {Calhoun}},
  \bibinfo {author} {\bibfnamefont {E.~L.}\ \bibnamefont {Read}}, \bibinfo
  {author} {\bibfnamefont {T.~K.}\ \bibnamefont {Ahn}}, \bibinfo {author}
  {\bibfnamefont {T.}~\bibnamefont {Man\v{c}al}}, \bibinfo {author}
  {\bibfnamefont {Y.~C.}\ \bibnamefont {Cheng}}, \bibinfo {author}
  {\bibfnamefont {R.~E.}\ \bibnamefont {Blankenship}},\ and\ \bibinfo {author}
  {\bibfnamefont {G.~R.}\ \bibnamefont {Fleming}},\ }\bibfield  {title}
  {\bibinfo {title} {Evidence for wavelike energy transfer through quantum
  coherence in photosynthetic systems},\ }\href@noop {} {\bibfield  {journal}
  {\bibinfo  {journal} {Nature}\ }\textbf {\bibinfo {volume} {446}},\ \bibinfo
  {pages} {782} (\bibinfo {year} {2007})}\BibitemShut {NoStop}%
\bibitem [{\citenamefont {Dorfman}\ \emph {et~al.}(2013)\citenamefont
  {Dorfman}, \citenamefont {Voronine}, \citenamefont {Mukamel},\ and\
  \citenamefont {Scully}}]{Dor132746}%
  \BibitemOpen
  \bibfield  {author} {\bibinfo {author} {\bibfnamefont {K.~E.}\ \bibnamefont
  {Dorfman}}, \bibinfo {author} {\bibfnamefont {D.~V.}\ \bibnamefont
  {Voronine}}, \bibinfo {author} {\bibfnamefont {S.}~\bibnamefont {Mukamel}},\
  and\ \bibinfo {author} {\bibfnamefont {M.~O.}\ \bibnamefont {Scully}},\
  }\bibfield  {title} {\bibinfo {title} {Photosynthetic reaction center as a
  quantum heat engine},\ }\href@noop {} {\bibfield  {journal} {\bibinfo
  {journal} {Proc. Natl. Acad. Sci.}\ }\textbf {\bibinfo {volume} {110}},\
  \bibinfo {pages} {2746} (\bibinfo {year} {2013})}\BibitemShut {NoStop}%
\bibitem [{\citenamefont {Creatore}\ \emph {et~al.}(2013)\citenamefont
  {Creatore}, \citenamefont {Parker}, \citenamefont {Emmott},\ and\
  \citenamefont {Chin}}]{Cre13253601}%
  \BibitemOpen
  \bibfield  {author} {\bibinfo {author} {\bibfnamefont {C.}~\bibnamefont
  {Creatore}}, \bibinfo {author} {\bibfnamefont {M.~A.}\ \bibnamefont
  {Parker}}, \bibinfo {author} {\bibfnamefont {S.}~\bibnamefont {Emmott}},\
  and\ \bibinfo {author} {\bibfnamefont {A.~W.}\ \bibnamefont {Chin}},\
  }\bibfield  {title} {\bibinfo {title} {Efficient biologically inspired
  photocell enhanced by delocalized quantum states},\ }\href@noop {} {\bibfield
   {journal} {\bibinfo  {journal} {Phys. Rev. Lett.}\ }\textbf {\bibinfo
  {volume} {111}},\ \bibinfo {pages} {253601} (\bibinfo {year}
  {2013})}\BibitemShut {NoStop}%
\bibitem [{\citenamefont {Kundu}\ \emph {et~al.}(2022)\citenamefont {Kundu},
  \citenamefont {Dani},\ and\ \citenamefont {Makri}}]{Kun22015101}%
  \BibitemOpen
  \bibfield  {author} {\bibinfo {author} {\bibfnamefont {S.}~\bibnamefont
  {Kundu}}, \bibinfo {author} {\bibfnamefont {R.}~\bibnamefont {Dani}},\ and\
  \bibinfo {author} {\bibfnamefont {N.}~\bibnamefont {Makri}},\ }\bibfield
  {title} {\bibinfo {title} {B800-to-b850 relaxation of excitation energy in
  bacterial light harvesting: All-state, all-mode path integral simulations},\
  }\href@noop {} {\bibfield  {journal} {\bibinfo  {journal} {J. Chem. Phys.}\
  }\textbf {\bibinfo {volume} {157}},\ \bibinfo {pages} {015101} (\bibinfo
  {year} {2022})}\BibitemShut {NoStop}%
\bibitem [{\citenamefont {Nakajima}(1958)}]{Nak58948}%
  \BibitemOpen
  \bibfield  {author} {\bibinfo {author} {\bibfnamefont {S.}~\bibnamefont
  {Nakajima}},\ }\bibfield  {title} {\bibinfo {title} {On quantum theory of
  transport phenomena -- steady diffusion},\ }\href@noop {} {\bibfield
  {journal} {\bibinfo  {journal} {Prog. Theor. Phys.}\ }\textbf {\bibinfo
  {volume} {20}},\ \bibinfo {pages} {948} (\bibinfo {year} {1958})}\BibitemShut
  {NoStop}%
\bibitem [{\citenamefont {Zwanzig}(1960)}]{Zwa601338}%
  \BibitemOpen
  \bibfield  {author} {\bibinfo {author} {\bibfnamefont {R.}~\bibnamefont
  {Zwanzig}},\ }\bibfield  {title} {\bibinfo {title} {Ensemble method in the
  theory of irreversibility},\ }\href@noop {} {\bibfield  {journal} {\bibinfo
  {journal} {J. Chem. Phys.}\ }\textbf {\bibinfo {volume} {33}},\ \bibinfo
  {pages} {1338} (\bibinfo {year} {1960})}\BibitemShut {NoStop}%
\bibitem [{\citenamefont {Redfield}(1957)}]{Red5719}%
  \BibitemOpen
  \bibfield  {author} {\bibinfo {author} {\bibfnamefont {A.~G.}\ \bibnamefont
  {Redfield}},\ }\href@noop {} {\bibfield  {journal} {\bibinfo  {journal} {IBM
  J.\ Res.\ Develop.}\ }\textbf {\bibinfo {volume} {1}},\ \bibinfo {pages} {19}
  (\bibinfo {year} {1957})}\BibitemShut {NoStop}%
\bibitem [{\citenamefont {Davies}(1974)}]{Dav7491}%
  \BibitemOpen
  \bibfield  {author} {\bibinfo {author} {\bibfnamefont {E.~B.}\ \bibnamefont
  {Davies}},\ }\bibfield  {title} {\bibinfo {title} {Markovian master
  equations},\ }\href@noop {} {\bibfield  {journal} {\bibinfo  {journal} {Comm.
  Math. Phys.}\ }\textbf {\bibinfo {volume} {39}},\ \bibinfo {pages} {91}
  (\bibinfo {year} {1974})}\BibitemShut {NoStop}%
\bibitem [{\citenamefont {Gorini}\ \emph {et~al.}(1976)\citenamefont {Gorini},
  \citenamefont {Kossakowski},\ and\ \citenamefont {Sudarshan}}]{Gor76821}%
  \BibitemOpen
  \bibfield  {author} {\bibinfo {author} {\bibfnamefont {V.}~\bibnamefont
  {Gorini}}, \bibinfo {author} {\bibfnamefont {A.}~\bibnamefont
  {Kossakowski}},\ and\ \bibinfo {author} {\bibfnamefont {E.~C.~G.}\
  \bibnamefont {Sudarshan}},\ }\bibfield  {title} {\bibinfo {title} {Completely
  positive dynamical semigroups of $n$-level systems},\ }\href@noop {}
  {\bibfield  {journal} {\bibinfo  {journal} {J. Math. Phys.}\ }\textbf
  {\bibinfo {volume} {17}},\ \bibinfo {pages} {821} (\bibinfo {year}
  {1976})}\BibitemShut {NoStop}%
\bibitem [{\citenamefont {Lindblad}(1976)}]{Lin76119}%
  \BibitemOpen
  \bibfield  {author} {\bibinfo {author} {\bibfnamefont {G.}~\bibnamefont
  {Lindblad}},\ }\bibfield  {title} {\bibinfo {title} {On the generators of
  quantum dynamical semigroups},\ }\href@noop {} {\bibfield  {journal}
  {\bibinfo  {journal} {Commun. Math. Phys.}\ }\textbf {\bibinfo {volume}
  {48}},\ \bibinfo {pages} {119} (\bibinfo {year} {1976})}\BibitemShut
  {NoStop}%
\bibitem [{\citenamefont {Breuer}\ \emph {et~al.}(2009)\citenamefont {Breuer},
  \citenamefont {Laine},\ and\ \citenamefont {Piilo}}]{Bre09210401}%
  \BibitemOpen
  \bibfield  {author} {\bibinfo {author} {\bibfnamefont {H.-P.}\ \bibnamefont
  {Breuer}}, \bibinfo {author} {\bibfnamefont {E.-M.}\ \bibnamefont {Laine}},\
  and\ \bibinfo {author} {\bibfnamefont {J.}~\bibnamefont {Piilo}},\ }\bibfield
   {title} {\bibinfo {title} {Measure for the degree of non-markovian behavior
  of quantum processes in open systems},\ }\href
  {https://doi.org/10.1103/PhysRevLett.103.210401} {\bibfield  {journal}
  {\bibinfo  {journal} {Phys. Rev. Lett.}\ }\textbf {\bibinfo {volume} {103}},\
  \bibinfo {pages} {210401} (\bibinfo {year} {2009})}\BibitemShut {NoStop}%
\bibitem [{\citenamefont {Rivas}\ \emph {et~al.}(2010)\citenamefont {Rivas},
  \citenamefont {Huelga},\ and\ \citenamefont {Plenio}}]{Riv10050403}%
  \BibitemOpen
  \bibfield  {author} {\bibinfo {author} {\bibfnamefont {{\'A}.}~\bibnamefont
  {Rivas}}, \bibinfo {author} {\bibfnamefont {S.~F.}\ \bibnamefont {Huelga}},\
  and\ \bibinfo {author} {\bibfnamefont {M.~B.}\ \bibnamefont {Plenio}},\
  }\bibfield  {title} {\bibinfo {title} {Entanglement and non-markovianity of
  quantum evolutions},\ }\href {https://doi.org/10.1103/PhysRevLett.105.050403}
  {\bibfield  {journal} {\bibinfo  {journal} {Phys. Rev. Lett.}\ }\textbf
  {\bibinfo {volume} {105}},\ \bibinfo {pages} {050403} (\bibinfo {year}
  {2010})}\BibitemShut {NoStop}%
\bibitem [{\citenamefont {Rivas}\ \emph {et~al.}(2014)\citenamefont {Rivas},
  \citenamefont {Huelga},\ and\ \citenamefont {Plenio}}]{Riv14094001}%
  \BibitemOpen
  \bibfield  {author} {\bibinfo {author} {\bibfnamefont {{\'A}.}~\bibnamefont
  {Rivas}}, \bibinfo {author} {\bibfnamefont {S.~F.}\ \bibnamefont {Huelga}},\
  and\ \bibinfo {author} {\bibfnamefont {M.~B.}\ \bibnamefont {Plenio}},\
  }\bibfield  {title} {\bibinfo {title} {Quantum non-markovianity:
  Characterization, quantification and detection},\ }\href
  {https://doi.org/10.1088/0034-4885/77/9/094001} {\bibfield  {journal}
  {\bibinfo  {journal} {Rep. Prog. Phys.}\ }\textbf {\bibinfo {volume} {77}},\
  \bibinfo {pages} {094001} (\bibinfo {year} {2014})}\BibitemShut {NoStop}%
\bibitem [{\citenamefont {Breuer}\ \emph {et~al.}(2016)\citenamefont {Breuer},
  \citenamefont {Laine}, \citenamefont {Piilo},\ and\ \citenamefont
  {Vacchini}}]{Bre16021002}%
  \BibitemOpen
  \bibfield  {author} {\bibinfo {author} {\bibfnamefont {H.-P.}\ \bibnamefont
  {Breuer}}, \bibinfo {author} {\bibfnamefont {E.-M.}\ \bibnamefont {Laine}},
  \bibinfo {author} {\bibfnamefont {J.}~\bibnamefont {Piilo}},\ and\ \bibinfo
  {author} {\bibfnamefont {B.}~\bibnamefont {Vacchini}},\ }\bibfield  {title}
  {\bibinfo {title} {Colloquium: Non-markovian dynamics in open quantum
  systems},\ }\href {https://doi.org/10.1103/RevModPhys.88.021002} {\bibfield
  {journal} {\bibinfo  {journal} {Rev. Mod. Phys.}\ }\textbf {\bibinfo {volume}
  {88}},\ \bibinfo {pages} {021002} (\bibinfo {year} {2016})}\BibitemShut
  {NoStop}%
\bibitem [{\citenamefont {de~Vega}\ and\ \citenamefont
  {Alonso}(2017)}]{Dev17015001}%
  \BibitemOpen
  \bibfield  {author} {\bibinfo {author} {\bibfnamefont {I.}~\bibnamefont
  {de~Vega}}\ and\ \bibinfo {author} {\bibfnamefont {D.}~\bibnamefont
  {Alonso}},\ }\bibfield  {title} {\bibinfo {title} {Dynamics of non-markovian
  open quantum systems},\ }\href {https://doi.org/10.1103/RevModPhys.89.015001}
  {\bibfield  {journal} {\bibinfo  {journal} {Rev. Mod. Phys.}\ }\textbf
  {\bibinfo {volume} {89}},\ \bibinfo {pages} {015001} (\bibinfo {year}
  {2017})}\BibitemShut {NoStop}%
\bibitem [{\citenamefont {Hornberger}(2009)}]{Hor09221}%
  \BibitemOpen
  \bibfield  {author} {\bibinfo {author} {\bibfnamefont {K.}~\bibnamefont
  {Hornberger}},\ }\bibfield  {title} {\bibinfo {title} {Introduction to
  decoherence theory},\ }in\ \href
  {https://doi.org/10.1007/978-3-540-88169-8_5} {\emph {\bibinfo {booktitle}
  {Entanglement and Decoherence: Foundations and Modern Trends}}},\ \bibinfo
  {series} {Lecture Notes in Physics}, Vol.\ \bibinfo {volume} {768},\ \bibinfo
  {editor} {edited by\ \bibinfo {editor} {\bibfnamefont {A.}~\bibnamefont
  {Buchleitner}}, \bibinfo {editor} {\bibfnamefont {C.}~\bibnamefont
  {Viviescas}},\ and\ \bibinfo {editor} {\bibfnamefont {M.}~\bibnamefont
  {Tiersch}}}\ (\bibinfo  {publisher} {Springer},\ \bibinfo {address} {Berlin,
  Heidelberg},\ \bibinfo {year} {2009})\ pp.\ \bibinfo {pages}
  {221--276}\BibitemShut {NoStop}%
\bibitem [{\citenamefont {Feynman}\ and\ \citenamefont {\mbox{Vernon,
  Jr.}}(1963)}]{Fey63118}%
  \BibitemOpen
  \bibfield  {author} {\bibinfo {author} {\bibfnamefont {R.~P.}\ \bibnamefont
  {Feynman}}\ and\ \bibinfo {author} {\bibfnamefont {F.~L.}\ \bibnamefont
  {\mbox{Vernon, Jr.}}},\ }\bibfield  {title} {\bibinfo {title} {The theory of
  a general quantum system interacting with a linear dissipative system},\
  }\href@noop {} {\bibfield  {journal} {\bibinfo  {journal} {Ann. Phys.}\
  }\textbf {\bibinfo {volume} {24}},\ \bibinfo {pages} {118} (\bibinfo {year}
  {1963})}\BibitemShut {NoStop}%
\bibitem [{\citenamefont {Caldeira}\ and\ \citenamefont
  {Leggett}(1981)}]{Cal81211}%
  \BibitemOpen
  \bibfield  {author} {\bibinfo {author} {\bibfnamefont {A.~O.}\ \bibnamefont
  {Caldeira}}\ and\ \bibinfo {author} {\bibfnamefont {A.~J.}\ \bibnamefont
  {Leggett}},\ }\bibfield  {title} {\bibinfo {title} {Influence of dissipation
  on quantum tunneling in macroscopic systems},\ }\href@noop {} {\bibfield
  {journal} {\bibinfo  {journal} {Phys. Rev. Lett.}\ }\textbf {\bibinfo
  {volume} {46}},\ \bibinfo {pages} {211} (\bibinfo {year} {1981})}\BibitemShut
  {NoStop}%
\bibitem [{\citenamefont {Caldeira}\ and\ \citenamefont
  {Leggett}(1983{\natexlab{a}})}]{Cal83587}%
  \BibitemOpen
  \bibfield  {author} {\bibinfo {author} {\bibfnamefont {A.~O.}\ \bibnamefont
  {Caldeira}}\ and\ \bibinfo {author} {\bibfnamefont {A.~J.}\ \bibnamefont
  {Leggett}},\ }\bibfield  {title} {\bibinfo {title} {Path integral approach to
  quantum brownian motion},\ }\href@noop {} {\bibfield  {journal} {\bibinfo
  {journal} {Physica A}\ }\textbf {\bibinfo {volume} {121}},\ \bibinfo {pages}
  {587} (\bibinfo {year} {1983}{\natexlab{a}})}\BibitemShut {NoStop}%
\bibitem [{\citenamefont {Caldeira}\ and\ \citenamefont
  {Leggett}(1983{\natexlab{b}})}]{Cal83374}%
  \BibitemOpen
  \bibfield  {author} {\bibinfo {author} {\bibfnamefont {A.~O.}\ \bibnamefont
  {Caldeira}}\ and\ \bibinfo {author} {\bibfnamefont {A.~J.}\ \bibnamefont
  {Leggett}},\ }\bibfield  {title} {\bibinfo {title} {Quantum tunnelling in a
  dissipative system},\ }\href@noop {} {\bibfield  {journal} {\bibinfo
  {journal} {Ann. Phys.}\ }\textbf {\bibinfo {volume} {149}},\ \bibinfo {pages}
  {374} (\bibinfo {year} {1983}{\natexlab{b}})},\ \bibinfo {note} {[Erratum:
  {\bf 153}, 445 (1984)]}\BibitemShut {NoStop}%
\bibitem [{\citenamefont {Haake}\ and\ \citenamefont
  {Reibold}(1985)}]{Haa852462}%
  \BibitemOpen
  \bibfield  {author} {\bibinfo {author} {\bibfnamefont {F.}~\bibnamefont
  {Haake}}\ and\ \bibinfo {author} {\bibfnamefont {R.}~\bibnamefont
  {Reibold}},\ }\bibfield  {title} {\bibinfo {title} {Strong damping and
  low-temperature anomalies for the harmonic oscillator},\ }\href@noop {}
  {\bibfield  {journal} {\bibinfo  {journal} {Phys. Rev. A}\ }\textbf {\bibinfo
  {volume} {32}},\ \bibinfo {pages} {2462} (\bibinfo {year}
  {1985})}\BibitemShut {NoStop}%
\bibitem [{\citenamefont {Leggett}\ \emph {et~al.}(1987)\citenamefont
  {Leggett}, \citenamefont {Chakravarty}, \citenamefont {Dorsey}, \citenamefont
  {Fisher}, \citenamefont {Garg},\ and\ \citenamefont {Zwerger}}]{Leg871}%
  \BibitemOpen
  \bibfield  {author} {\bibinfo {author} {\bibfnamefont {A.~J.}\ \bibnamefont
  {Leggett}}, \bibinfo {author} {\bibfnamefont {S.}~\bibnamefont
  {Chakravarty}}, \bibinfo {author} {\bibfnamefont {A.~T.}\ \bibnamefont
  {Dorsey}}, \bibinfo {author} {\bibfnamefont {M.~P.~A.}\ \bibnamefont
  {Fisher}}, \bibinfo {author} {\bibfnamefont {A.}~\bibnamefont {Garg}},\ and\
  \bibinfo {author} {\bibfnamefont {W.}~\bibnamefont {Zwerger}},\ }\bibfield
  {title} {\bibinfo {title} {Dynamics of the dissipative two-state system},\
  }\href@noop {} {\bibfield  {journal} {\bibinfo  {journal} {Rev. Mod. Phys.}\
  }\textbf {\bibinfo {volume} {59}},\ \bibinfo {pages} {1} (\bibinfo {year}
  {1987})},\ \bibinfo {note} {{\bf 67}, 725-726 (Erratum) (1995)}\BibitemShut
  {NoStop}%
\bibitem [{\citenamefont {Grabert}\ \emph {et~al.}(1988)\citenamefont
  {Grabert}, \citenamefont {Schramm},\ and\ \citenamefont {Ingold}}]{Gra88115}%
  \BibitemOpen
  \bibfield  {author} {\bibinfo {author} {\bibfnamefont {H.}~\bibnamefont
  {Grabert}}, \bibinfo {author} {\bibfnamefont {P.}~\bibnamefont {Schramm}},\
  and\ \bibinfo {author} {\bibfnamefont {G.~L.}\ \bibnamefont {Ingold}},\
  }\bibfield  {title} {\bibinfo {title} {Quantum brownian motion: The
  functional integral approach},\ }\href@noop {} {\bibfield  {journal}
  {\bibinfo  {journal} {Phys. Rep.}\ }\textbf {\bibinfo {volume} {168}},\
  \bibinfo {pages} {115} (\bibinfo {year} {1988})}\BibitemShut {NoStop}%
\bibitem [{\citenamefont {H{\"a}nggi}\ and\ \citenamefont
  {Ingold}(2005)}]{Han05026106}%
  \BibitemOpen
  \bibfield  {author} {\bibinfo {author} {\bibfnamefont {P.}~\bibnamefont
  {H{\"a}nggi}}\ and\ \bibinfo {author} {\bibfnamefont {G.-L.}\ \bibnamefont
  {Ingold}},\ }\bibfield  {title} {\bibinfo {title} {Fundamental aspects of
  quantum brownian motion},\ }\href {https://doi.org/10.1063/1.1853631}
  {\bibfield  {journal} {\bibinfo  {journal} {Chaos}\ }\textbf {\bibinfo
  {volume} {15}},\ \bibinfo {pages} {026105} (\bibinfo {year}
  {2005})}\BibitemShut {NoStop}%
\bibitem [{\citenamefont {Xu}\ \emph {et~al.}(2026)\citenamefont {Xu},
  \citenamefont {Vadimov}, \citenamefont {Stockburger},\ and\ \citenamefont
  {Ankerhold}}]{Xu26xxx}%
  \BibitemOpen
  \bibfield  {author} {\bibinfo {author} {\bibfnamefont {M.}~\bibnamefont
  {Xu}}, \bibinfo {author} {\bibfnamefont {V.}~\bibnamefont {Vadimov}},
  \bibinfo {author} {\bibfnamefont {J.~T.}\ \bibnamefont {Stockburger}},\ and\
  \bibinfo {author} {\bibfnamefont {J.}~\bibnamefont {Ankerhold}},\ }\bibfield
  {title} {\bibinfo {title} {Colloquium: Simulating non-markovian dynamics in
  open quantum systems},\ }\href {https://doi.org/10.1103/w3nw-hbjc} {\bibfield
   {journal} {\bibinfo  {journal} {Rev. Mod. Phys.}\ ,\ } (\bibinfo {year}
  {2026})}\BibitemShut {NoStop}%
\bibitem [{\citenamefont {Dirac}(1964)}]{Dir64}%
  \BibitemOpen
  \bibfield  {author} {\bibinfo {author} {\bibfnamefont {P.~A.~M.}\
  \bibnamefont {Dirac}},\ }\href@noop {} {\emph {\bibinfo {title} {Lectures on
  Quantum Mechanics}}}\ (\bibinfo  {publisher} {Belfer Graduate School of
  Science, Yeshiva University},\ \bibinfo {address} {New York},\ \bibinfo
  {year} {1964})\BibitemShut {NoStop}%
\bibitem [{\citenamefont {Sundermeyer}(1982)}]{Sun82}%
  \BibitemOpen
  \bibfield  {author} {\bibinfo {author} {\bibfnamefont {K.}~\bibnamefont
  {Sundermeyer}},\ }\href@noop {} {\emph {\bibinfo {title} {Constrained
  Dynamics: With Applications to Yang--Mills Theory, General Relativity,
  Classical Spin, Dual String Model}}},\ \bibinfo {series} {Lecture Notes in
  Physics}, Vol.\ \bibinfo {volume} {169}\ (\bibinfo  {publisher}
  {Springer-Verlag},\ \bibinfo {address} {Berlin},\ \bibinfo {year}
  {1982})\BibitemShut {NoStop}%
\bibitem [{\citenamefont {Henneaux}\ and\ \citenamefont
  {Teitelboim}(1992)}]{Hen92}%
  \BibitemOpen
  \bibfield  {author} {\bibinfo {author} {\bibfnamefont {M.}~\bibnamefont
  {Henneaux}}\ and\ \bibinfo {author} {\bibfnamefont {C.}~\bibnamefont
  {Teitelboim}},\ }\href@noop {} {\emph {\bibinfo {title} {Quantization of
  Gauge Systems}}}\ (\bibinfo  {publisher} {Princeton University Press},\
  \bibinfo {address} {Princeton, NJ},\ \bibinfo {year} {1992})\BibitemShut
  {NoStop}%
\bibitem [{\citenamefont {Gitman}\ and\ \citenamefont {Tyutin}(1990)}]{Git90}%
  \BibitemOpen
  \bibfield  {author} {\bibinfo {author} {\bibfnamefont {D.~M.}\ \bibnamefont
  {Gitman}}\ and\ \bibinfo {author} {\bibfnamefont {I.~V.}\ \bibnamefont
  {Tyutin}},\ }\href {https://doi.org/10.1007/978-3-642-83938-2} {\emph
  {\bibinfo {title} {Quantization of Fields with Constraints}}},\ Springer
  Series in Nuclear and Particle Physics\ (\bibinfo  {publisher}
  {Springer-Verlag},\ \bibinfo {address} {Berlin},\ \bibinfo {year}
  {1990})\BibitemShut {NoStop}%
\bibitem [{\citenamefont {Faddeev}\ and\ \citenamefont
  {Jackiw}(1988)}]{Fad881692}%
  \BibitemOpen
  \bibfield  {author} {\bibinfo {author} {\bibfnamefont {L.}~\bibnamefont
  {Faddeev}}\ and\ \bibinfo {author} {\bibfnamefont {R.}~\bibnamefont
  {Jackiw}},\ }\bibfield  {title} {\bibinfo {title} {Hamiltonian reduction of
  unconstrained and constrained systems},\ }\href
  {https://doi.org/10.1103/PhysRevLett.60.1692} {\bibfield  {journal} {\bibinfo
   {journal} {Phys. Rev. Lett.}\ }\textbf {\bibinfo {volume} {60}},\ \bibinfo
  {pages} {1692} (\bibinfo {year} {1988})}\BibitemShut {NoStop}%
\bibitem [{\citenamefont {Weinberg}(2015)}]{Wei15}%
  \BibitemOpen
  \bibfield  {author} {\bibinfo {author} {\bibfnamefont {S.}~\bibnamefont
  {Weinberg}},\ }\href@noop {} {\emph {\bibinfo {title} {Lectures on Quantum
  Mechanics}}}\ (\bibinfo  {publisher} {Cambridge University Press},\ \bibinfo
  {year} {2015})\BibitemShut {NoStop}%
\bibitem [{\citenamefont {Senjanovic}(1976)}]{Sen76227}%
  \BibitemOpen
  \bibfield  {author} {\bibinfo {author} {\bibfnamefont {P.}~\bibnamefont
  {Senjanovic}},\ }\bibfield  {title} {\bibinfo {title} {Path integral
  quantization of field theories with second-class constraints},\ }\href
  {https://doi.org/10.1016/0003-4916(76)90062-2} {\bibfield  {journal}
  {\bibinfo  {journal} {Ann. Phys.}\ }\textbf {\bibinfo {volume} {100}},\
  \bibinfo {pages} {227} (\bibinfo {year} {1976})}\BibitemShut {NoStop}%
\bibitem [{\citenamefont {Shao}(2004)}]{Sha045053}%
  \BibitemOpen
  \bibfield  {author} {\bibinfo {author} {\bibfnamefont {J.~S.}\ \bibnamefont
  {Shao}},\ }\bibfield  {title} {\bibinfo {title} {Decoupling quantum
  dissipation interaction via stochastic fields},\ }\href@noop {} {\bibfield
  {journal} {\bibinfo  {journal} {J. Chem. Phys.}\ }\textbf {\bibinfo {volume}
  {120}},\ \bibinfo {pages} {5053} (\bibinfo {year} {2004})}\BibitemShut
  {NoStop}%
\bibitem [{\citenamefont {Yan}\ \emph {et~al.}(2004)\citenamefont {Yan},
  \citenamefont {Yang}, \citenamefont {Liu},\ and\ \citenamefont
  {Shao}}]{Yan04216}%
  \BibitemOpen
  \bibfield  {author} {\bibinfo {author} {\bibfnamefont {Y.~A.}\ \bibnamefont
  {Yan}}, \bibinfo {author} {\bibfnamefont {F.}~\bibnamefont {Yang}}, \bibinfo
  {author} {\bibfnamefont {Y.}~\bibnamefont {Liu}},\ and\ \bibinfo {author}
  {\bibfnamefont {J.~S.}\ \bibnamefont {Shao}},\ }\bibfield  {title} {\bibinfo
  {title} {Hierarchical approach based on stochastic decoupling to dissipative
  systems},\ }\href@noop {} {\bibfield  {journal} {\bibinfo  {journal} {Chem.
  Phys. Lett.}\ }\textbf {\bibinfo {volume} {395}},\ \bibinfo {pages} {216}
  (\bibinfo {year} {2004})}\BibitemShut {NoStop}%
\bibitem [{\citenamefont {Fukui}(1981)}]{Fuk81363}%
  \BibitemOpen
  \bibfield  {author} {\bibinfo {author} {\bibfnamefont {K.}~\bibnamefont
  {Fukui}},\ }\bibfield  {title} {\bibinfo {title} {The path of chemical
  reactions --- the irc approach},\ }\href
  {https://doi.org/10.1021/ar00072a001} {\bibfield  {journal} {\bibinfo
  {journal} {Accounts of Chemical Research}\ }\textbf {\bibinfo {volume}
  {14}},\ \bibinfo {pages} {363} (\bibinfo {year} {1981})}\BibitemShut
  {NoStop}%
\bibitem [{\citenamefont {Miller}\ \emph {et~al.}(1980)\citenamefont {Miller},
  \citenamefont {Handy},\ and\ \citenamefont {Adams}}]{Mil8099}%
  \BibitemOpen
  \bibfield  {author} {\bibinfo {author} {\bibfnamefont {W.~H.}\ \bibnamefont
  {Miller}}, \bibinfo {author} {\bibfnamefont {N.~C.}\ \bibnamefont {Handy}},\
  and\ \bibinfo {author} {\bibfnamefont {J.~E.}\ \bibnamefont {Adams}},\
  }\bibfield  {title} {\bibinfo {title} {Reaction path hamiltonian for
  polyatomic molecules},\ }\href@noop {} {\bibfield  {journal} {\bibinfo
  {journal} {J. Chem. Phys.}\ }\textbf {\bibinfo {volume} {72}},\ \bibinfo
  {pages} {99} (\bibinfo {year} {1980})}\BibitemShut {NoStop}%
\end{thebibliography}

%

\end{document}